\documentclass[twocolumn,showpacs,preprintnumbers,superscriptaddress,aps]{revtex4-1}

\usepackage{graphicx,color}

\usepackage{dcolumn}

\usepackage{bm}
\usepackage{here}
\usepackage[]{lineno}
\usepackage{mathrsfs} 

\begin{document}

\preprint{}

\title{Large spontaneous Hall effect arising from collinear antiferromagnetism in Ce$_2$PtGe$_6$}

\author{Hayata Matsuda}
\thanks{These authors contributed equally to this work.}
\affiliation{Department of Physics, Kobe University, Kobe, Hyogo 657-8501, Japan}

\author{Ruo Hibino}
\thanks{These authors contributed equally to this work.}
\affiliation{Department of Physics, Kobe University, Kobe, Hyogo 657-8501, Japan}

\author{Chihiro Tabata}
\affiliation{Materials Sciences Research Center, Japan Atomic Energy Agency, Tokai 319-1195, Japan}
\affiliation{Advanced Science Research Center, Japan Atomic Energy Agency, Tokai 319-1195, Japan}

\author{Koji Kaneko}
\affiliation{Materials Sciences Research Center, Japan Atomic Energy Agency, Tokai 319-1195, Japan}
\affiliation{Advanced Science Research Center, Japan Atomic Energy Agency, Tokai 319-1195, Japan}

\author{Nonoka Higa}
\affiliation{Graduate School of Advanced Science and Engineering, Hiroshima University, Higashi-Hiroshima 739-8526, Japan}

\author{Takahiro Onimaru}
\affiliation{Graduate School of Advanced Science and Engineering, Hiroshima University, Higashi-Hiroshima 739-8526, Japan}

\author{Hiroto Tanaka}
\affiliation{Department of Physics, Kobe University, Kobe, Hyogo 657-8501, Japan}

\author{Hideki Tou}
\affiliation{Department of Physics, Kobe University, Kobe, Hyogo 657-8501, Japan}

\author{Hitoshi Sugawara}
\affiliation{Department of Physics, Kobe University, Kobe, Hyogo 657-8501, Japan}

\author{Junichi Hayashi}
\affiliation{Muroran Institute of Technology, Muroran, Hokkaido 050-8585, Japan}

\author{Keiki Takeda}
\affiliation{Muroran Institute of Technology, Muroran, Hokkaido 050-8585, Japan}

\author{Hisashi Kotegawa}
\affiliation{Department of Physics, Kobe University, Kobe, Hyogo 657-8501, Japan}

\date{\today}

\begin{abstract}
The spontaneous Hall effect, corresponding to a zero-field anomalous Hall effect (AHE), is induced by symmetry breaking associated with ferromagnetism.
Studies in recent years, however, have revealed that antiferromagnetic (AFM) states characterized by magnetic point groups that allow ferromagnetism can also break the relevant symmetries and induce AHE without a large net magnetization. 
Here, we report that the AFM system Ce$_2$PtGe$_6$ exhibits a pronounced spontaneous Hall effect. 
Single-crystal neutron scattering experiments demonstrate that Ce$_2$PtGe$_6$ exhibits a collinear AFM structure with a propagation vector $\bm{q}=0$.
The small net magnetization of $\sim 10^{-3}$ $\mu_B$/Ce indicates that the observed AHE arises from symmetry breaking inherent to its AFM structure.
The anomalous Hall conductivity (AHC) reaches $300$ $\Omega^{-1}$cm$^{-1}$, which exceeds the intrinsic AHC of related compounds such as Ce$_2$CuGe$_6$ and Ce$_2$PdGe$_6$. 
This large AHC, most likely attributed to the large spin-orbit coupling of the Pt atoms, provides a platform for understanding the interplay between the Berry curvatures and localized $f$-moments with an AFM configuration. 
\end{abstract}

\maketitle

\section{Introduction}

The emergence of an anomalous Hall effect is governed by the symmetry of the magnetically ordered state, specifically its magnetic point group, regardless of whether the state is ferromagnetic (FM) or antiferromagnetic (AFM) \cite{Chen14,Suzuki17,Smejkal20}.
Among the 122 magnetic point groups, 31 can represent FM or ferrimagnetic states with significant spontaneous magnetization.
The same magnetic point groups can also describe certain AFM states, which can host nonzero spontaneous magnetization through the Dzyaloshinskii--Moriya (DM) interaction \cite{DM}. 
If the system is conductive, the emergence of AHE is allowed, and the anomalous Hall conductivity (AHC) is determined by the Berry curvature in momentum space.
Some AFM systems meeting this criterion, such as Mn$_3$Sn and Mn$_3$Ge, have been reported to exhibit large AHCs comparable to those of ferromagnets \cite{Nakatsuji2015,Kiyohara16,Nayak16}. 
Other materials showing large zero-field AHCs ($>100$ $\Omega^{-1}$cm$^{-1}$) include $\alpha-$Mn under pressure, Mn$_3$Sb, NbMnP, and TaMnP 
\cite{Akiba20,Hayashi,Kotegawa_NbMnP,Kotegawa_TaMnP}.

Recently, zero-field AHE has also been discovered in the $f$-electron compounds Ce$_2$CuGe$_6$ and Ce$_2$PdGe$_6$ via a symmetry-based approach \cite{Kotegawa_216}. 
The intrinsic AHC, derived from Berry curvature, is estimated to be 80--100 $\Omega^{-1}$cm$^{-1}$ for both compounds. 
In high-quality Ce$_2$CuGe$_6$ crystals, extrinsic contributions dominate the AHC at low temperatures. 
This mechanism is also allowed by its magnetic symmetry.
These AFM $f$-electron systems provide an ideal platform for exploring Berry curvature physics arising from the interplay between conduction electrons and localized moment, such as the Kondo effect.

Ce$_2$CuGe$_6$ crystallize in the orthorhombic $Cmce$ structure \cite{Qi}.
Earlier reports assigned many $R$$_2$$T$Ge$_6$ ($R$: rare earth element, $T$: transition metal) compounds to the $Amm2$ space group \cite{Sologub}, but more recent studies have corrected this to $Cmce$ \cite{Qi,Penc}.
Ce$_2$CuGe$_6$ undergoes an AFM transition at $T_{\rm N} = 15$~K with a weak FM moment along the $c$-axis \cite{Sologub,Nakashima}. 
Its magnetic structure is a $\bm{q}=0$ AFM configuration belonging to the magnetic point group $m'm'm$, which permits a spontaneous magnetization along the $c$ axis \cite{Qi}. 
Although the magnetic structure of Ce$_2$PdGe$_6$ ($T_{\rm N}=11.2$ K) remains undetermined, its AHE and similar magnetic behavior suggest the same symmetry \cite{Kotegawa_216}. 
While related compounds with $T$ = Ag, Ni, Pt have been reported \cite{Sologub}, no single crystals had previously been synthesized.
In this study, we grew single crystals of Ce$_2$PtGe$_6$, which is expected to exhibit strong spin-orbit coupling originating from the Pt-$5d$ electrons, and we investigated its magnetic and transport properties.

\begin{table}[b]
\caption{Structural parameters of Ce$_2$PtGe$_6$ and Ce$_2$CuGe$_6$, determined by single-crystal X-ray diffraction measurements at $T=293\, \mathrm{K}$. Ce$_2$PtGe$_6$ was grown by the self-flux method, while Ce$_2$CuGe$_6$ was obtained by the Bi-flux method.}

\label{t1}
\vspace{1ex}
\begin{center}
\begin{tabular}{lcc}\hline
Temperature & 293 K & 293 K \\
Formula & Ce$_2$PtGe$_6$ & Ce$_2$CuGe$_6$ \\
Crystal system & orthorhombic & orthorhombic \\
Space group & $Cmce$ (no.64) & $Cmce$ (no.64) \\
$a$ (\AA) & 8.2921(6) & 8.4070(2) \\
$b$ (\AA) & 8.1565(5) & 8.1281(2) \\
$c$ (\AA) & 22.1344(14) & 21.5305(7) \\
$V$ (\AA$^3$) & 1497.05(17) & 1471.24(7) \\
$Z$ & 8 & 8 \\
Independent reflections & 1158 & 1218 \\
Residual factor $R1$ & 0.0336 & 0.0243 \\
$wR2$ & 0.0756 & 0.0536 \\
\hline
\end{tabular}
\\
\vspace{2ex}
\begin{tabular}{ccccccc}\hline
Ce$_2$PtGe$_6$ \\
& Wyckoff & & & & & \\
atom & position & $x$ & $y$ & $z$ & $U_{eq}$ (\AA$^2$) & occup. \\

\hline
Ce & 16g & 0.25047 & 0.37432 & 0.08213 & 0.0055 & 1 \\
Pt & 8f & 0 & 0.12481 & 0.14420 & 0.0057 & 0.988 \\
Ge1 & 16g & 0.27580 & 0.12440 & 0.19464 & 0.0083 & 1 \\
Ge2 & 8f & 0 & 0.12180 & 0.46318 & 0.0074 & 1 \\
Ge3 & 8f & 0 & 0.12830 & 0.03038 & 0.0071 & 1 \\
Ge4 & 8f & 0 & 0.34768 & 0.30534 & 0.0088 & 1 \\
Ge5 & 8f & 0 & 0.40216 & 0.19474 & 0.0078 & 1 \\
\hline
\end{tabular}
\\
\vspace{2ex}
\begin{tabular}{ccccccc}\hline
Ce$_2$CuGe$_6$ \\
& Wyckoff & & & & & \\
atom & position & $x$ & $y$ & $z$ & $U_{eq}$ (\AA$^2$) & occup. \\
\hline
Ce & 16g & 0.25076 & 0.37488 & 0.08328 & 0.0054 & 1 \\
Cu & 8f & 0 & 0.12461 & 0.14687 & 0.0087 & 0.981 \\
Ge1 & 16g & 0.27773 & 0.12525 & 0.19387 & 0.0079 & 1 \\
Ge2 & 8f & 0 & 0.12231 & 0.46160 & 0.0073 & 1 \\
Ge3 & 8f & 0 & 0.12829 & 0.03221 & 0.0065 & 1 \\
Ge4 & 8f & 0 & 0.34426 & 0.30623 & 0.0080 & 1 \\
Ge5 & 8f & 0 & 0.40565 & 0.19387 & 0.0080 & 1 \\
\hline
\end{tabular}
\end{center}
\end{table}

\begin{figure}[b]
\includegraphics[width=0.9\linewidth]{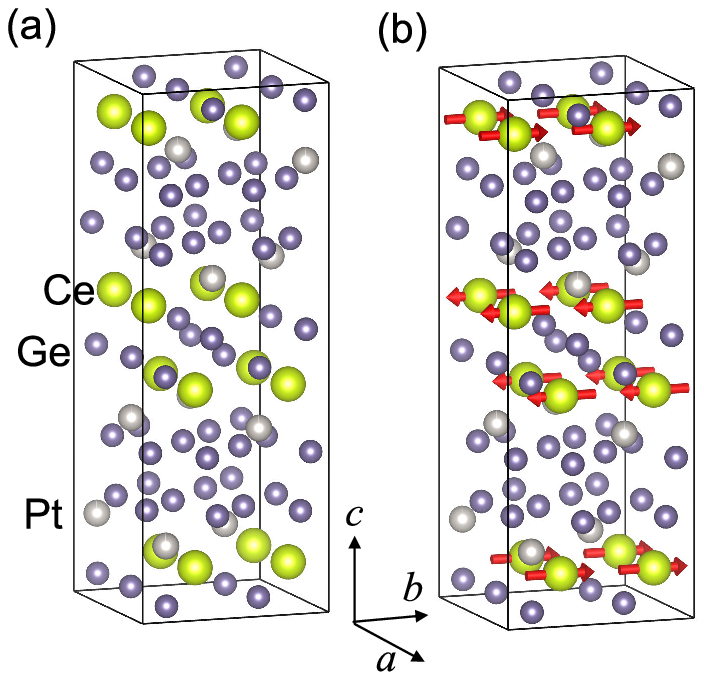}
\caption{(a) Orthorhombic crystal structure of Ce$_2$PtGe$_6$. The space group was determined to be $Cmce$. (b) Magnetic structure of Ce$_2$PtGe$_6$ determined in this study. The propagation vector is $\bm{q}=0$, and the magnetic point group is represented by $m'm'm$, which allows AHE. This structure consists of the large collinear AFM component along the $b$-axis ($1.7$ $\mu_B$) and the negligibly small FM component along the $c$-axis ($0.001$ $\mu_B$).}
\label{fig1}
\end{figure}

\begin{figure*}[ht]
\includegraphics[width=\linewidth]{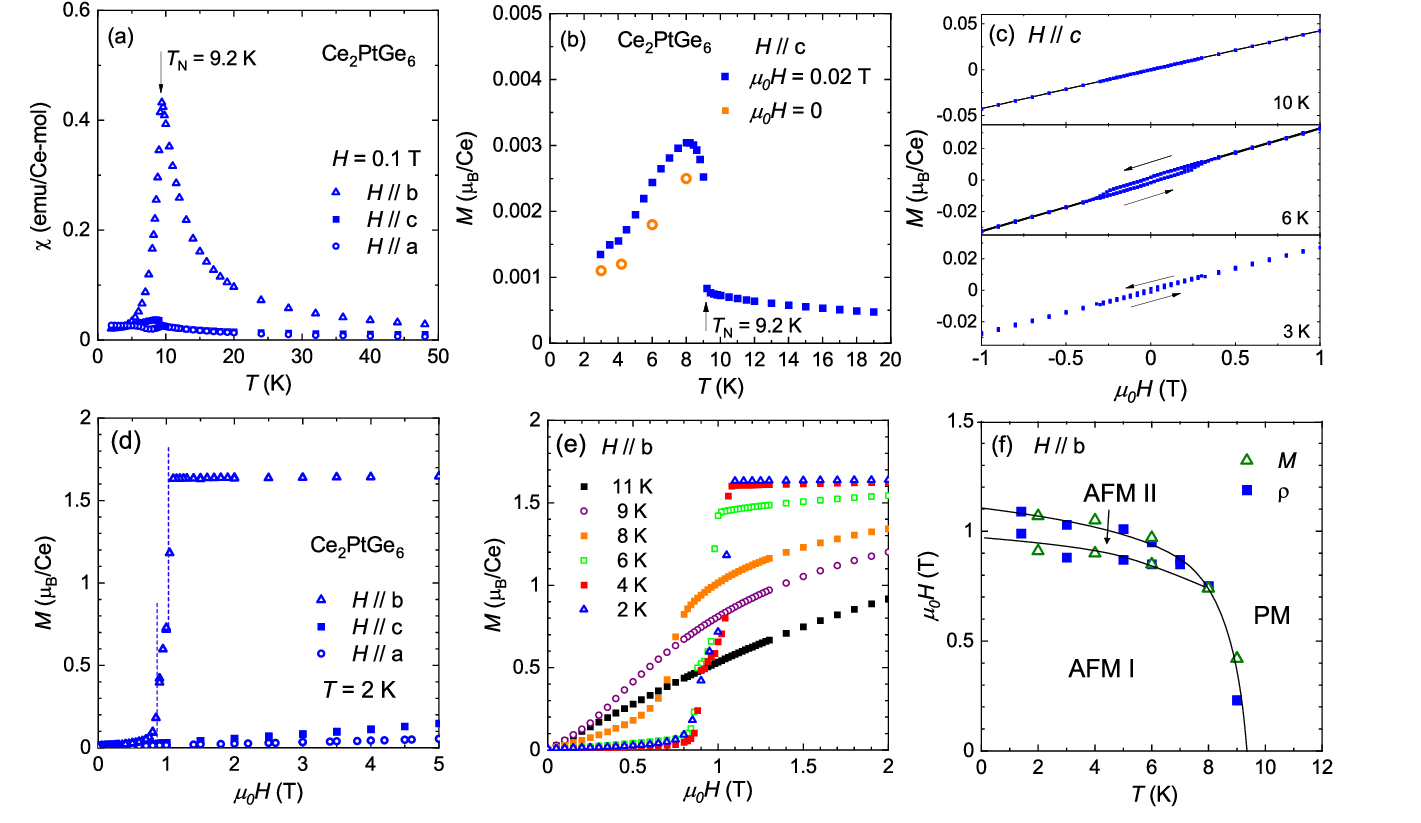}
\caption{(a) Temperature dependence of magnetic susceptibility for $H \parallel a$, $b$, and $c$. The clear kink at $T_{\rm N}=9.2$ K for $H \parallel b$ indicates the AFM transition. (b) Temperature dependence of magnetization for $H \parallel c$ at 0.02 T. The increase in magnetization below $T_{\rm N}$ suggests that a small FM component is involved in the ordered state. The spontaneous magnetizations obtained from the hysteresis loop are also plotted (orange circles). The magnetization decreases towards low temperatures. (c) Field dependence of magnetization at several temperatures. In the ordered state, weak spontaneous magnetization with hysteresis appears. The coercive force increases at low temperatures, and reaches almost 5 T at 4.2 K. At 3 K, the maximum field of the experimental setup was not sufficient to reverse the magnetic domain. The data were taken after the field-cooling in both positive and negative fields. The spontaneous magnetization clearly decreases with decreasing temperature. (d) The field dependence of the magnetization at 2 K. The metamagnetic transitions were observed at approximately 1 T for $H \parallel b$. (e) The temperature dependences of the metamagnetic transition for $H \parallel b$. The plotted data were obtained during the field-decreasing process. They show small hysteresis with widths of less than 0.05 T. (f) The magnetic phase diagram for $H \parallel b$. The AFM I corresponds to zero-field AFM state. The AFM II phase emerges in the narrow field region.}
\label{m}
\end{figure*}

\section{Experimental procedure}

Single crystals of Ce$_2$PtGe$_6$ were grown by the self-flux method. 
In contrast to Ce$_2T$Ge$_6$ ($T=$ Cu, Pd, and Au) \cite{Nakashima,Kotegawa_216}, the Bi-flux method was unsuccessful.
A molar mixture of ${\rm Ce} : {\rm Pt} : {\rm Ge} = 10 : 3 : 50$ was placed in an alumina crucible, sealed in an evacuated quartz tube, and heated to 1050$^\circ$C over 6 hours. 
After holding at this temperature for 6 hours, the mixture was slowly cooled to 850$^\circ$C at a rate of 5$^\circ$C/h. 
Excess flux was removed by centrifugation. 
Crystal orientations were determined by Laue measurements, and plate-like single crystals with a large $c$-plane were obtained.
Magnetization measurements were performed using a Magnetic Property Measurement System (MPMS). 
We performed specific heat measurement using a CRYOGENIC CFMS-9T.
Electrical resistivity and Hall measurements were carried out using a four-probe method. 
The Hall resistivity was symmetrized with respect to magnetic field to eliminate contact misalignment effects.
Crystal symmetry and lattice parameters were determined by single-crystal X-ray diffraction measurements using a Rigaku Saturn724
Diffractometer with a multilayer mirror monochromated Mo-K$_{\alpha}$ radiation at room temperature.
The chemical compositions of the crystals were checked by X-ray fluorescence (XRF) using a JEOL JSX-1000S.

Single-crystal neutron diffraction measurements were carried out on the thermal triple-axis spectrometer TAS-1 at the research reactor, JRR-3, Tokai, Japan. The spectrometer was operated in the triple-axis mode with a fixed neutron wavelength of 2.36 \AA. A collimation sequence of open-80'-80'-80' was employed, together with a pyrolytic graphite (PG) filter placed before the sample to suppress higher-order contamination. Two plate-like thin crystals (total mass: $\sim$ 1.6 mg) were co-aligned and attached to an aluminum plate with GE varnish to have the horizontal $\left(H\ 0\ L\ \right)$ scattering plane. The sample was cooled down to $\sim$ 1.7 K using a top-loading $\rm{^4He}$ cryostat \cite{kaneko}. Initial neutron diffraction measurements were performed in the unpolarized mode using the PG monochromator and analyzer to determine the magnetic propagation vector and collect integrated intensity data. Subsequently, polarized neutron measurements were carried out using a double-focusing Heusler monochromator and analyzer. Uniaxial polarization analysis was performed using a Helmholtz coil in combination with a Mezei-type flipper.

\section{Experimental results and discussion}

\subsection{Crystal structure}

Table I lists the structural parameters of Ce$_2$PtGe$_6$ and Ce$_2$CuGe$_6$ determined by single-crystal X-ray diffraction measurements.
Both crystals had been reported as a space group $Amm2$ \cite{Sologub,Strydom}; however, Ce$_2$CuGe$_6$ was subsequently corrected by $Cmce$ \cite{Qi}.
Our measurements suggest that Ce$_2$PtGe$_6$ also crystallizes in the same $Cmce$ structure.
The volume of Ce$_2$PtGe$_6$ is 1.8\% larger than that of Ce$_2$CuGe$_6$.
A small deficiency at the transition-metal site was observed, while other sites were sufficiently occupied and the occupancies were fixed to 1. 
The crystal structure of Ce$_2$PtGe$_6$ is shown in Fig.~1(a), which was is illustrated using {\it VESTA} \cite{VESTA}.
The chemical compositions investigated by X-ray fluorescence (XRF) were Ce$_{2.08}$Pt$_{0.94}$Ge$_{5.98}$ and Ce$_{1.98}$Cu$_{1.06}$Ge$_{5.96}$.
In this measurement, the deficiency at the transition-metal site was confirmed only in Ce$_2$PtGe$_6$.

\subsection{Magnetization measurements}

Figure~\ref{m}(a) shows the temperature dependence of the magnetic susceptibility for Ce$_2$PtGe$_6$ under magnetic fields applied along the three axes.
A pronounced magnetic anisotropy with the easy $b$-axis was observed, similar to Ce$_2$CuGe$_6$ and Ce$_2$PdGe$_6$ \cite{Yaguchi,Kotegawa_216}.
A kink at $T_{\rm N} = 9.2$~K appears for $H \parallel b$, indicating the AFM transition.
Figure~\ref{m}(b) shows magnetization for $\mu_0H =0.02$ ${\rm T} \parallel c$, which increases below $T_{\rm N}$, indicating a weak FM component in the ordered state. 
Figure~\ref{m}(c) demonstrates that clear hysteresis with spontaneous magnetization emerges below $T_{\rm N}$. 
This behavior is consistent with an $m'm'm$ AFM structure with a DM-interaction induced $c$-axis magnetization, as in Ce$_2$CuGe$_6$ \cite{Qi}.
Figure~\ref{m}(b) also shows the temperature variation of the spontaneous magnetization (orange circles), which obviously decreases with decreasing temperature, as well as the magnetization at 0.02 T. 
Such behavior was not seen in $T=$ Cu, and not clear in $T=$ Pd.
It is conjectured to arise from the competition between the AFM interaction, which favors an antiparallel configuration, and the DM interaction.
The spontaneous magnetization was $1\times10^{-3}~\mu_B$/Ce at low temperatures. 
Magnetizations up to 5 T along the three crystal axes, measured at 2 K, are shown in Fig.~\ref{m}(d).
Two metamagnetic transitions appeared at approximately 1 T for $H \parallel b$, similar to Ce$_2$CuGe$_6$ and Ce$_2$PdGe$_6$ \cite{Yaguchi}.
The transition fields are lower than those for Ce$_2$CuGe$_6$ ($\sim1.1$ and $\sim1.4$ T) and Ce$_2$PdGe$_6$ ($\sim1.0$ and $\sim1.2$ T), probably due to the difference in $T_{\rm N}$.
The magnetic moment above the metamagnetic transition reaches $\sim1.65$ $\mu_B$/Ce, which is comparable to those of Ce$_2$CuGe$_6$ and Ce$_2$PdGe$_6$ \cite{Yaguchi}.
Figure \ref{m}(e) shows the temperature dependences of the metamagnetic transitions.
Two transitions are confirmed at low temperatures, changing into one transition above 8 K. 
The phase diagram for $H \parallel b$ is shown in Fig. ~\ref{m}(f).

\subsection{Specific heat measurement}

Figure 3 shows the temperature dependence of specific heat of Ce$_2$PtGe$_6$ in the form of $C/T$.
A clear peak was observed at $T_{\rm N}$, which was consistent with that determined in the magnetization measurements. 
The $C/T$ continues to decrease even at the lowest temperature of 2 K, where $C/T$ was approximately 50 mJ mol$^{-1}$ K$^{-2}$.
This suggests that the electronic specific heat coefficient of Ce$_2$PtGe$_6$ is much less than 50 mJ mol$^{-1}$ K$^{-2}$.
The small $C/T$ above $T_{\rm N}$ indicates that the phonon contribution in $C/T$ is not significant in this temperature region. 
The calculated entropy is also shown in the figure.
The entropy at $T_{\rm N}$ is approximately 90 \% of $R\ln2$, which is similar to that in Ce$_2$CuGe$_6$ \cite{Nakashima}. 
These indicate that the crystal-filed ground state of the $4f$ electron is doublet and the localized character is strong for Ce$_2$PtGe$_6$, similar to Ce$_2$CuGe$_6$.

\begin{figure}[ht]
\includegraphics[width=0.9\linewidth]{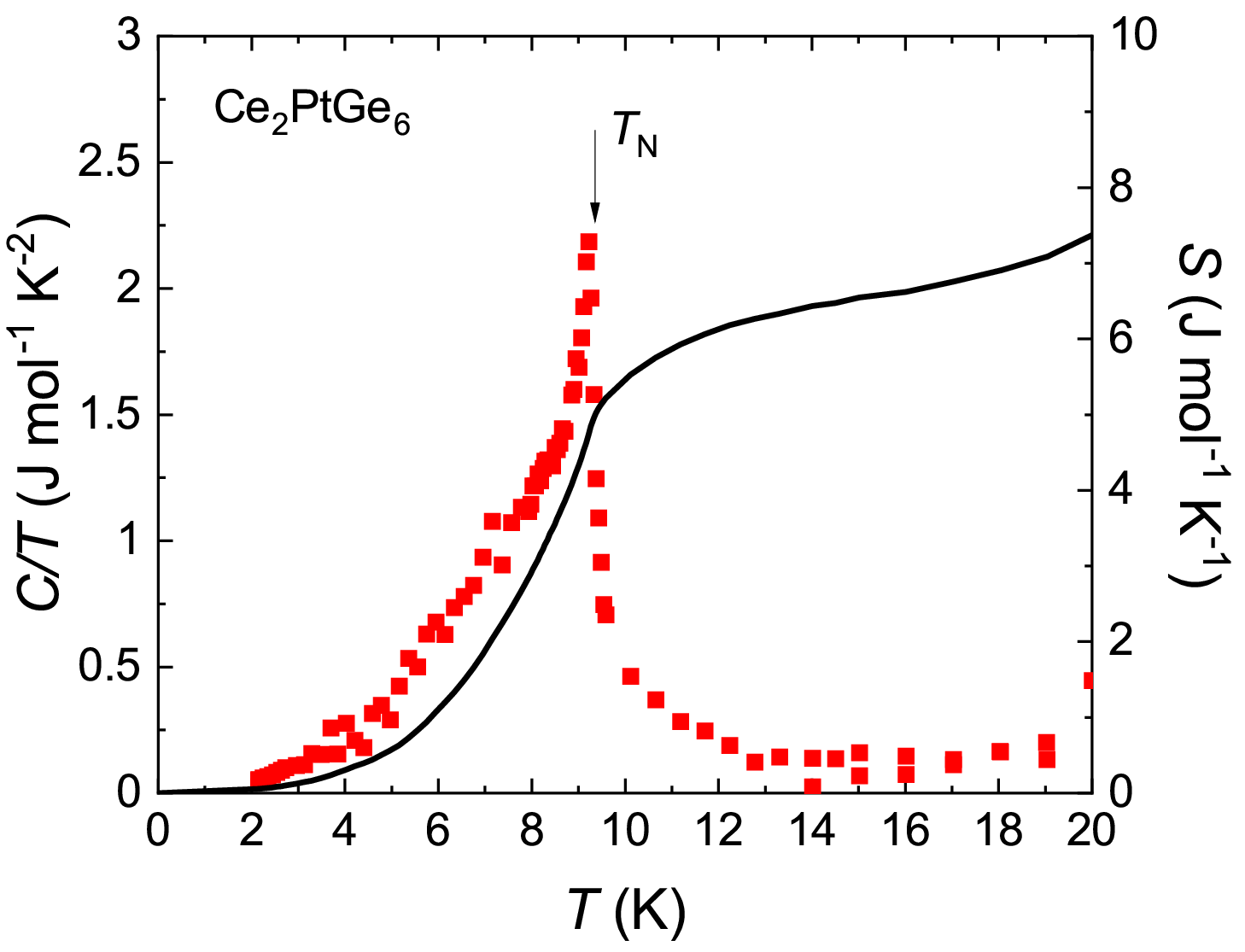}
\caption{Temperature dependence of the specific heat per Ce mole of Ce$_2$PtGe$_6$ in the form of $C/T$. The entropy at $T_{\rm N}$ was estimated to be approximately 5 J mol$^{-1}$ K$^{-1}$, which is 90 \% of $R\ln2$.}
\label{C}
\end{figure}

\subsection{Neutron diffraction measurements}
\begin{figure*}[ht]
\includegraphics[width=\linewidth]{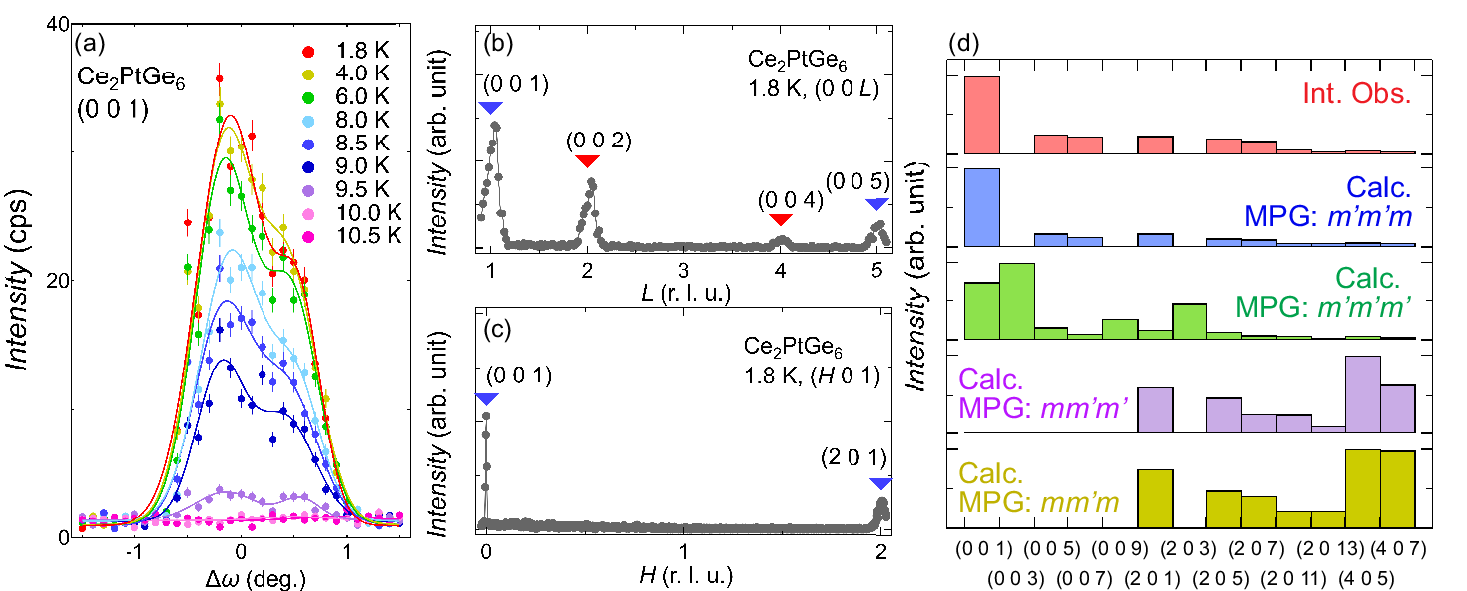}
\caption{(a) Temperature dependence of peak profiles of the magnetic reflection $\left(0\ 0\ 1\right)$. Solid curves show the two-gaussian fitting curves. Line scan along the (b) $\left(0\ 0\ L\right)$ and (c) $\left(H\ 0\ 1\right)$ at 1.8 K. Red arrows represent nuclear reflections and blue ones represent magnetic reflections in the panels (b) and (c). (d) Intensities of magnetic reflections accessible in the present experimental setup. Red data represent the observed intensities, while data in other colors represent the calculated intensities based on symmetrically independent magnetic structures with the given magnetic point groups (MPG). All data are normalized to the strongest intensity in each model.}
\label{fig_4_ND}
\end{figure*}
Single-crystal neutron diffraction measurements were performed to determine the magnetic structure of Ce$_2$PtGe$_6$ in the ordered phase. First, we measured unpolarized neutron diffraction using PG monochromator and analyzer and observed that several reflections develop below $T_{\rm N}$. The $\left(0\ 0\ 1\right)$ reflection, one of the prominent reflections shown in Fig. \ref{fig_4_ND}(a), is forbidden by the extinction rules of the crystal structure. This clearly indicates that the observed peak originates from the magnetic order. The asymmetric peak shape in the omega scan is attributed to a slight misalignment between co-aligned sample crystals. 
Furthermore, we carried out line scans along the $\left(0\ 0\ L\right)$ and $\left(H\ 0\ 1\right)$ as shown in Figs. \ref{fig_4_ND}(b) and (c), respectively. No additional magnetic peaks were detected except for $\left(0\ 0\ L\right)$ ($L$: odd) and $\left(H\ 0\ 1\right)$ ($H$: even). These observations reveal that the propagation vector of the AFM order is $\bm{q} = 0$, which is consistent with that of Ce$_2$CuGe$_6$\cite{Qi}.

\begin{figure}[ht]
\includegraphics[width=0.9\linewidth]{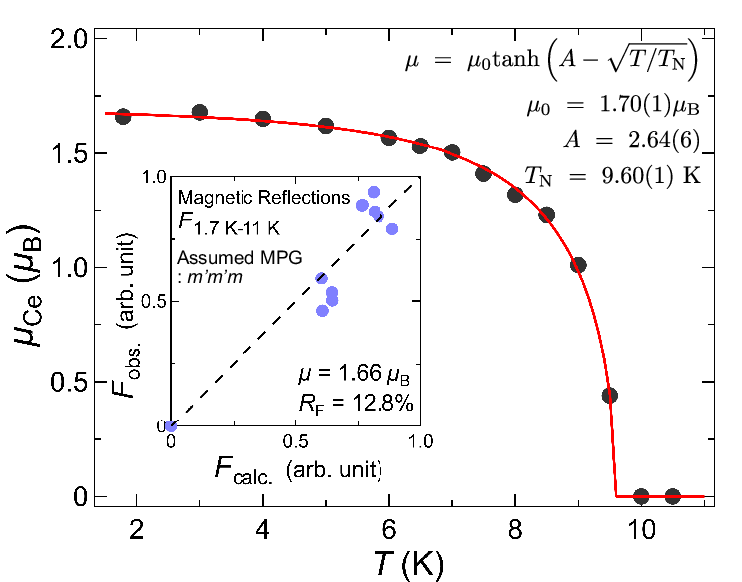}
\caption{Temperature dependence of the ordered magnetic moment estimated from the integrated intensities of the $\left(0\ 0\ 1\right)$. The inset shows the $F_{\rm o}$--$F_{\rm c}$ plot assuming the magnetic point group $m'm'm$. The red solid curve shows a fitting result based on the phenomenological equation, $\mu = \mu_{0}{\rm tanh}\left(A-\sqrt{T/T_{\rm N}}\right)$ to reproduce the observed magnetic moment $\mu_{\rm Ce}$.}
\label{fig_5_ND}
\end{figure}

Based on the group-theoretical analysis using the parent space group $Cmce$ and the propagation vector $\bm{q} = 0$ in Ce$_2$PtGe$_6$, eight symmetrically allowed magnetic structure models $\left(\rm{A_{g} \oplus A_{u} \oplus B_{1g} \oplus B_{1u} \oplus B_{2g} \oplus B_{2u} \oplus B_{3g} \oplus B_{3u}}\right)$ are obtained. Four models among them yield no finite magnetic reflection intensities under the conditions of $\left(0\ 0\ L\right)$ ($L$: odd) and $\left(H\ 0\ 1\right)$ ($H$: even). Therefore, we compared the observed magnetic reflection intensities with calculated ones for the remaining four magnetic structure models, as summarized in Fig. \ref{fig_4_ND}(d).
It is assumed that the ordered moments align along the $b$-axis in the calculation, which is indicated by our polarized neutron diffraction measurements (Details are given below). 
The comparison clearly demonstrates that the calculation for the magnetic point group $m'm'm$ (corresponding to the $\rm{B_{1g}}$ representation) satisfactorily reproduces the experimental data. Note that the resulting structure with $m'm'm$ for Ce$_2$PtGe$_6$ is the same structure as that of Ce$_2$CuGe$_6$ \cite{Qi}.
Finally, we performed a least-squares fitting to quantitatively reproduce the observed magnetic reflection intensities. The inset of Fig. \ref{fig_5_ND} represents the $F_{\rm obs.}$--$F_{\rm calc.}$ plot based on the magnetic structure with $m'm'm$. In the analysis, we used structural parameters determined from our single-crystal X-ray measurements (shown in Table. \ref{t1}), and only the magnitude of the ordered magnetic moment was treated as a free parameter. The optimized ordered moment was 1.66 $\mu_{\rm B}$ at 1.8 K, resulting in $R_F\ =\ 12.8$\%. This ordered moment is comparable to the saturated moment for $H \parallel b$ shown in Fig.~2(d). Figure \ref{fig_5_ND} shows the temperature dependence of the ordered magnetic moment estimated from the integrated intensities of the $\left(0\ 0\ 1\right)$ reflection.


\begin{figure}[ht]
\includegraphics[width=1.0\linewidth]{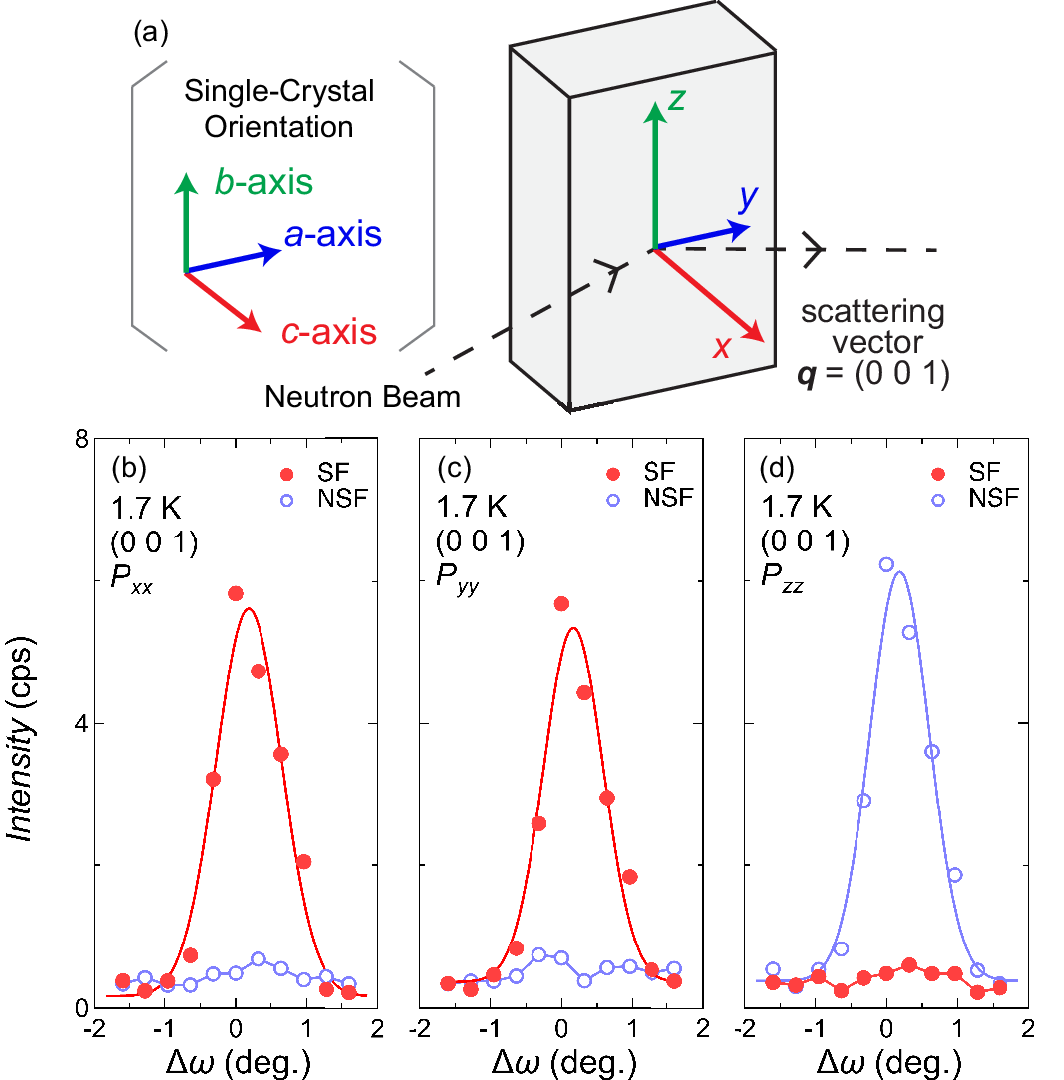}
\caption{(a) Schematic geometry of the polarized neutron diffraction measurement of the (0 0 1) reflection. Peak profiles for the $\left(0\ 0\ 1\right)$ reflection at 1.7 K measured in the polarized neutron channels (b) $P_{xx}$, (c) $P_{yy}$, and (d) $P_{zz}$. Red and blue symbols correspond to the SF and NSF signals, respectively.}
\label{fig_6_ND}
\end{figure}

Furthermore, we carried out polarized neutron diffraction measurements using a Heusler monochromator and analyzer to identify the moment orientation precisely. In the present measurements, the scattering plane was set perpendicular to the $b$-axis. Polarized neutron diffraction enables selective probing of the orientation of magnetic moments by manipulating the neutron spins with respect to the scattering vector $\bm{q}$. The relationship between the signal channels and the detectable component of the magnetic moment is summarized in Table \ref{t2}. A spin flip (SF) signal and a non-spin-flip (NSF) signal were switched using a spin flipper positioned before the sample. 
Figure \ref{fig_6_ND} shows the peak profiles of the $\left(0\ 0\ 1\right)$ reflection measured at 1.7 K. Since the scattering vector is parallel to the $c$-axis, the $x$ direction in Table \ref{t2} corresponds to the crystallographic $c$-axis. Based on the definition in Table \ref{t2} and Fig. \ref{fig_6_ND}(a), the $y$ and $z$ directions correspond to the crystallographic $a$- and $b$-axes, respectively. 
In the $P_{xx}$ and $P_{yy}$ modes, the $\left(0\ 0\ 1\right)$ reflection intensities were equally observed in the SF channel whereas no peak was found in the NSF channel as shown in Figs. \ref{fig_6_ND}(b) and (c). This clearly indicates that the peak is of purely magnetic origin and the ordered magnetic moments have no $M_y$ component, i.e., no $a$-axis component. This conclusion is supported by the opposite polarization dependence observed in the $P_{zz}$ mode (Fig. \ref{fig_6_ND}(d)).
In addition, we also observed the same polarization dependence for another magnetic reflection $\left(2\ 0\ 1\right)$, suggesting that the ordered moment has no component along the $c$-axis as well (data not shown). 
Note that the spontaneous magnetization of $0.001$ $\mu_B$ along the $c$-axis observed in the magnetization measurements is less than the resolution of the present experiments.
Taken together, the polarized neutron diffraction unambiguously identified that the magnetic moment of the Ce ion in the ordered phase is along $b$-axis (collinear AFM along the $b$-axis with the magnetic point group $m'm'm$) within the present experimental accuracy.

\begin{table}[htb]
\caption{General relationship between the types of the signals and their origins inducing a finite scattering cross section. $M_i$ denotes the component of the magnetic moment along each direction. Here, the $z$ direction is defined as perpendicular to the scattering plane, whereas the $x$ direction is parallel to the scattering vector.}
\label{t2}
\begin{center}
\setlength{\tabcolsep}{12pt}
\begin{tabular}{c|ccc}
&$P_{xx}$&$P_{yy}$&$P_{zz}$\\
\hline
SF& $M_y$, $M_z$ & $M_z$&$M_y$ \\
\hline
NSF& Nuclear & Nuclear, $M_y$ & Nuclear, $M_z$\\

\end{tabular}
\end{center}
\end{table}

\subsection{Transport measurements}

\begin{figure}[ht]
\includegraphics[width=0.7\linewidth]{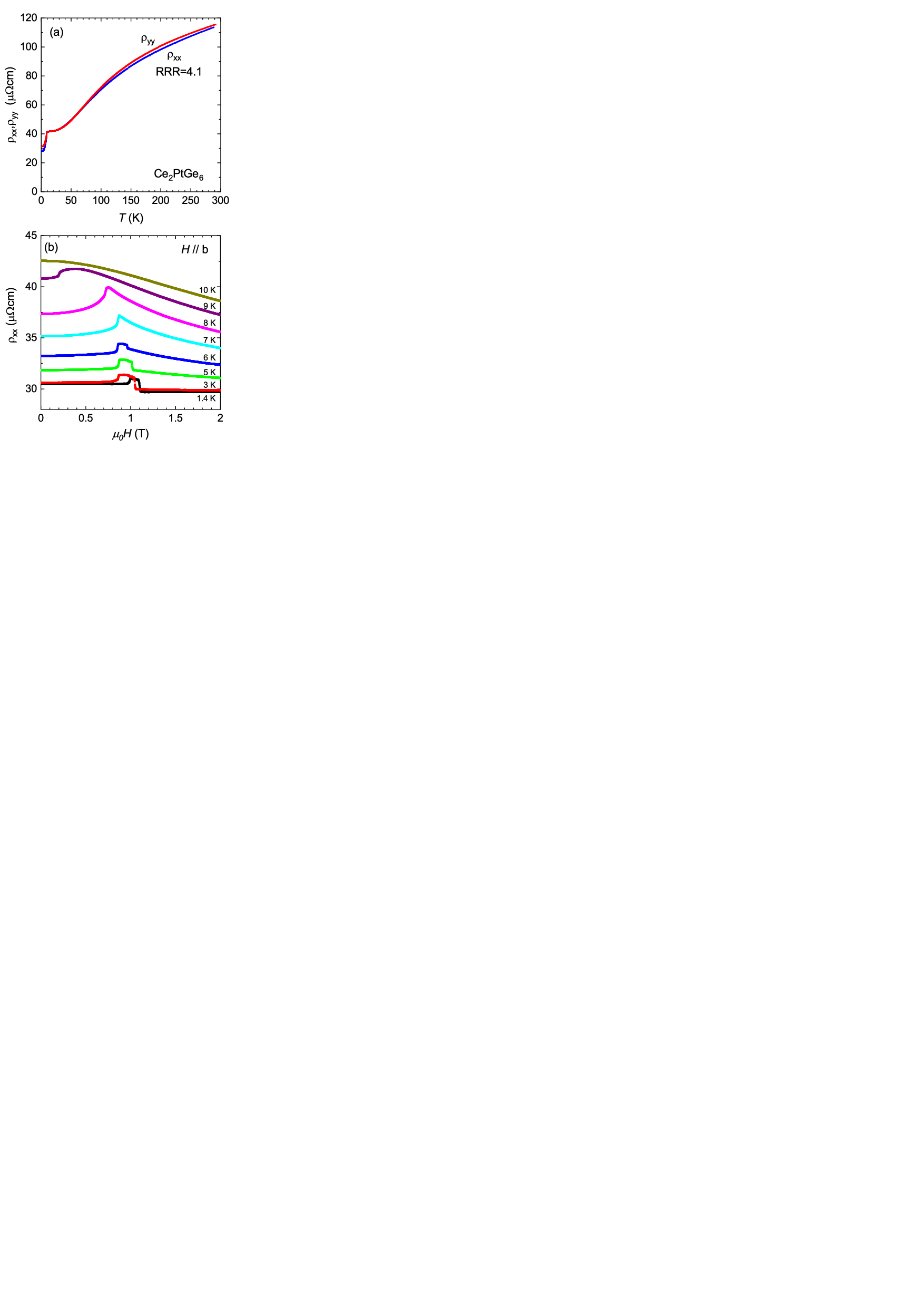}
\caption{(a) Temperature dependences of electric resistivity. Anisotropy between $\rho_{xx}$ and $\rho_{yy}$ was weak. (b) The magnetic field dependence of $\rho_{xx}$ for $H \parallel b$. The metamagnetic transitions induce the abrupt anomaly in $\rho_{xx}$. }
\label{fig7}
\end{figure}

Figure~\ref{fig7}(a) shows the temperature dependences of the electrical resistivity, $\rho_{xx}$ and $\rho_{yy}$, for Ce$_2$PtGe$_6$.
Here, $x$, $y$, and $z$ axes correspond to $a$, $b$, and $c$ directions, respectively.
Overall temperature dependence is similar to that of the polycrystalline sample \cite{Strydom}, but the residual resistivity ratio (RRR) decreased from 7.6 (polycrystal) to 4.1.
It is presumed that the present single crystals have a deviation from the stoichiometric ratio, because it was necessary to deviate the starting composition from the stoichiometric ratio in the self-flux method.
The broad shoulder in $\rho$ is likely attributed to the crystal-electric-field excitation, and anisotropy between $\rho_{xx}$ and $\rho_{yy}$ was quite weak.
Figure~\ref{fig7}(b) shows the field dependence of $\rho_{xx}$ under $H \parallel b$.
Abrupt changes in $\rho_{xx}$ appear corresponding to the metamagnetic transitions, as in the magnetization.
The data are plotted in the magnetic phase diagram for $H \parallel b$ shown in Fig.~\ref{m}(f).

\begin{figure*}[ht]
\includegraphics[width=0.9\linewidth]{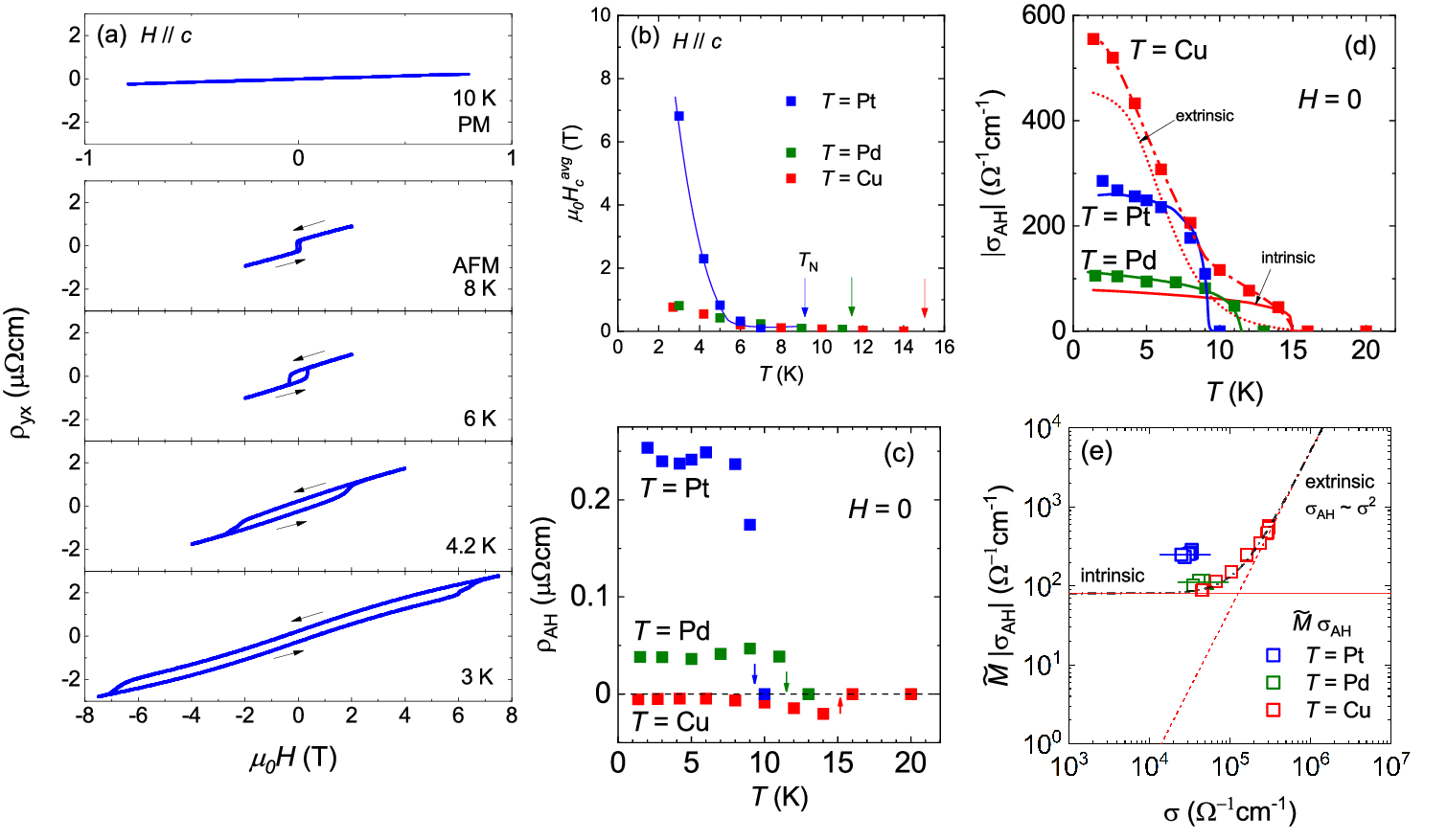}
\caption{(a) Field dependence of $\rho_{yx}$ of Ce$_2$PtGe$_6$ for $H \parallel c$. The clear hysteresis appears below $T_{\rm N}$ and nonzero $\rho_{yx}$ remains at zero field. (b) Temperature dependence of the coercive field, $H_c$ for Ce$_2T$Ge$_6$ ($T=$ Cu, Pd, and Pt). The $H_c$ for $T=$ Pt significantly increases with decreasing temperature in contrast to other two compounds. (c) Temperature dependence of the anomalous Hall resistivity, $\rho_{\rm AH}$, for Ce$_2T$Ge$_6$ ($T=$ Pt, Pd, and Cu). The sign of $\rho_{\rm AH}$ was defined as that obtained when the positive magnetic field was reduced to zero. The largest $|\rho_{\rm AH}|$ was obtained for $T=$ Pt.
(d) Temperature dependence of the AHC for three compounds. In high-quality Ce$_2$CuGe$_6$, both intrinsic and extrinsic contribution make the two-step temperature dependence \cite{Kotegawa_216}. $\sigma_{\rm AH}\simeq300$ $\Omega^{-1}$cm$^{-1}$ was obtained for Ce$_2$PtGe$_6$. The solid lines indicate $\sigma_{\rm AH} \sim M_{\rm AF}(T)$, which are assumed as the intrinsic contribution. (e) The scaling relation of the $\widetilde{M} \cdot |\sigma_{\rm AH}|$ versus $\sigma$, where $\widetilde{M}=M_{\rm AF}(0)/M_{\rm AF}(T)$.
}
\label{fig8}
\end{figure*}

Figure~\ref{fig8}(a) presents the magnetic-field dependence of Hall resistivity, $\rho_{yx}$.
Here, the magnetic fields were applied along the $c$-axis to switch the magnetic domain. 
No notable anomaly was seen in the paramagnetic (PM) state, while clear hysteresis appeared below $T_{\rm N}$.
The coercive field, $H_{c}$, significantly increases with decreasing temperature, and reaches $\sim7$ T at 3 K.
Figure ~\ref{fig8}(b) shows the temperature dependence of $H_c$ for Ce$_2T$Ge$_6$ ($T=$ Pt, Cu, and Pd).
Here, $H_c$ was estimated from the average value of the positive and negative fields, because it slightly shows an exchange-bias effect. 
At low temperatures, the $H_c$ for $T=$ Pt is much higher than the values of $\sim1$ T for $T=$ Cu and Pd \cite{Kotegawa_216}.
The spontaneous magnetization is $3\times10^{-2}~\mu_{\rm B}$/Ce for $T=$ Cu, while $1\times10^{-3}~\mu_B$/Ce for $T=$ Pd and Pt.
As an indicator of sample quality, RRR of $T=$ Cu is 20, whereas RRR of $T=$ Pd and Pt is 4.
Since both values are similar for $T=$ Pd and Pt, the large $H_c$ for $T=$ Pt at low temperatures cannot be explained by only the small spontaneous magnetization and the sample quality.

Figure~\ref{fig8}(c) shows the temperature dependence of anomalous Hall resistivity $\rho_{\rm AH}$, which corresponds to $\rho_{yx}$ at zero field.
In addition to Ce$_2$PtGe$_6$, the data for Ce$_2T$Ge$_6$ ($T=$ Cu and Pd) are also shown for comparison \cite{Kotegawa_216}.
Here, the sign of $\rho_{\rm AH}$ was determined from that obtained when the positive magnetic field was reduced to zero.
Obviously, the magnitude of $\rho_{\rm AH}$ for Ce$_2$PtGe$_6$ is the largest.
The sign of $\rho_{\rm AH}$ is positive of $T=$ Pt and Pd, while the sign is negative for $T=$ Cu, probably reflecting the sign of the Berry curvatures summed over the momentum space.

Figure~\ref{fig8}(d) shows the temperature dependence of the anomalous Hall conductivity $\sigma_{\rm AH}$, which was obtained through $\sigma_{\rm AH}=\sigma_{xy} \simeq \rho_{yx}/\rho_{xx}\rho_{yy}$ at zero fields.
Only Ce$_2$CuGe$_6$ shows a two-step increase, suggesting combined intrinsic and extrinsic contributions due to its high quality (${\rm RRR}=20$) \cite{Kotegawa_216}. 
For mechanisms of AHE, the dissipationless AHC, that is, $\sigma_{\rm AH} \propto \sigma^0$, is expected for the intrinsic contribution derived from the Berry curvature, while extrinsic contribution originating in the impurity scattering gives dissipative AHC; $\sigma_{\rm AH} \propto \sigma^2$ \cite{Tian}, where $\sigma=1/\rho$ is the electrical conductivity.
Therefore, the extrinsic contribution tends to appear in the low scattering regime \cite{Onoda2008}.
To simplify the problem, we assume $\sigma_{\rm AH} \propto M_{\rm AF}(T) \cdot \sigma^n$, where $M_{\rm AF}(T)$ indicates the temperature variation of the order parameter.
Figure~\ref{fig8}(e) shows the scaling relation, the $\widetilde{M} \cdot |\sigma_{\rm AH}|$ versus $\sigma$ is plotted, where $\widetilde{M}=M_{\rm AF}(0)/M_{\rm AF}(T)$.
$M_{\rm AF}(T)$ was obtained from the neutron scattering data reported for $T=$ Cu \cite{Qi} and obtained in this study for $T=$ Pt.
For $T=$ Pd, $M_{\rm AF}(T)$ of $T=$ Cu was used after normalization by $T_{\rm N}$. 
The extrinsic contribution, which gives $\sigma_{\rm AH} \propto \sigma^2$, appears for Ce$_2$CuGe$_6$ in high $\sigma$ region at low temperatures \cite{Kotegawa_216}, whereas such high $\sigma$ region is not realized in $T=$ Pd and $T=$ Pt.
For $T=$ Cu, the intrinsic contribution was estimated to be 80 $\Omega^{-1}$cm$^{-1}$, which was indicated by the solid red line in Fig. \ref{fig8}(e).
This intrinsic contribution is shown by the red solid curve for Ce$_2$CuGe$_6$ in Fig~\ref{fig8}(d) as $\sigma_{\rm AH}^{int}(T) \sim M_{\rm AF}(T)$.
The dissipationless intrinsic contribution can appear even just below $T_{\rm N}$, even though the resistivity is still high.
Towards low temperatures, the scattering is suppressed, then the dissipative extrinsic contribution is enhanced (the solid red curve).
For $T=$ Pt and Pd systems, the solid lines are also given as $\sigma_{\rm AH}^{int}(T) \sim M_{\rm AF}(T)$. 
The small increase in $\sigma_{\rm AH}$ for $T=$ Pt at low temperatures might indicate the presence of the extrinsic contribution, but the overall good scaling between the experimental data and the solid curves suggests that $\sigma_{\rm AH}$s for $T=$ Pd and Pt are dominated by the intrinsic contribution.
For $T=$ Cu and Pd, the intrinsic AHC is estimated to be 80--100 $\Omega^{-1}$cm$^{-1}$ at low temperature, while it reaches $250-300$ $\Omega^{-1}$cm$^{-1}$ for Ce$_2$PtGe$_6$.
This large value is comparable to that in Ce-based ferromagnetic Weyl semimetal CeAlSi \cite{Yang,Sakhya}.
It is an interesting issue to reveal what induces the large AHC in Ce$_2$PtGe$_6$ and how the Pt-$5d$ electrons contribute to it.
Especially, it is challenging for $f$-electron systems whether theoretical calculation can reproduce the AHC including the difference in the sign of the Hall response between $T=$ Cu and $T=$ Pd and Pt.

\section{Conclusion}

In summary, we have synthesized single crystals of Ce$_2$PtGe$_6$ and determined the collinear AFM structure using the polarized and unpolarized neutron diffraction measurements.
The symmetry of the obtained magnetic structure and the small net magnetization of $1\times10^{-3}~\mu_B$/Ce suggested that the AFM structure induces symmetry breaking associated with ferromagnetism.
We also observed a large spontaneous Hall effect originating from its AFM structure. 
The anomalous Hall conductivity reaches 300 $\Omega^{-1}$cm$^{-1}$, comparable to that of ferromagnets, and is most likely attributed to the intrinsic Berry curvature enhanced by the strong spin-orbit coupling of Pt. 
Another notable aspect observed in Ce$_2$PtGe$_6$ is a large coercive force. 
The spontaneous magnetization of $1\times10^{-3}~\mu_B$/Ce is similar between $T=$ Pt and Pd, whereas the coercive fields are about ten times different.
These findings not only shed light on Berry-curvature-induced transport phenomena in localized $f$-electron systems, but also provide a good platform for deepening the understanding of the stability and controllability of AFM domains under the strong spin-orbit coupling.

\section*{Acknowledgments}
This work was supported by JSPS KAKENHI Grant Nos. 21H04987, 23H04866, 23H04867, 23H04870, 23H04871, 25K00947, 25K23359, Iketani Science and Technology Foundation, and Murata Science Foundation.
The neutron experiments were performed at JRR-3 under Proposal Nos. I1289 and 2026A-A34.


\begin{thebibliography}{}
\bibitem{Chen14}
H. Chen, Q. Niu, A. H. MacDonald, Anomalous Hall effect arising from noncollinear antiferromagnetism, Phys. Rev. Lett. {\bf 112}, 017205 (2014).
\bibitem{Suzuki17}
M.-T. Suzuki, T. Koretsune, M. Ochi, and R. Arita,  Cluster multipole theory for anomalous Hall effect in antiferromagnets, Phys. Rev. B {\bf 95}, 094406 (2017).
\bibitem{Smejkal20}
L. \v{S}mejkal, R. Gon\'{a}zlez-Hern\'{a}ndez, T. Jungwirth, J. Sinova, Crystal time-reversal symmetry breaking and spontaneous Hall effect in collinear antiferromagnets, Sci. Adv. {\bf 6}, eaaz8809 (2020), and the supplementary material. 
\bibitem{DM}
I. Dzyaloshinsky, A thermodynamic theory of ``weak'' ferromagnetism of antiferromagnetics, J. Phys. Chem. Solids {\bf 4}, 241 (1958).
\bibitem{Nakatsuji2015}
S. Nakatsuji, N. Kiyohara, and T. Higo,  Large anomalous Hall effect in a non-collinear antiferromagnet at room temperature, Nature {\bf 527}, 212 (2015).
\bibitem{Kiyohara16}
N. Kiyohara, T. Tomita, and S. Nakatsuji, Giant anomalous Hall effect in the chiral antiferromagnet Mn$_3$Ge, Phys. Rev. Appli. {\bf 5}, 064009 (2016).
\bibitem{Nayak16}
A. K. Nayak, J. E. Fischer, Y. Sun, B. Yan, J. Karel, A. C. Komarek, C. Shekhar, N. Kumar, W. Schnelle, J. Kubler, C. Felser, and S. S. Parkin, Large anomalous Hall effect driven by a nonvanishing Berry curvature in the noncolinear antiferromagnet Mn$_3$Ge, Sci. Adv. {\bf 2}, e1501870 (2016).
\bibitem{Akiba20}
K. Akiba, K. Iwamoto, T. Sato, S. Araki, and T. C. Kobayashi, Anomalous Hall effect triggered by pressure-induced magnetic phase transition in $\alpha-$Mn, Phys. Rev. Reser. {\bf 2}, 043090 (2020).
\bibitem{Hayashi}
H. Hayashi, Y. Shirako, L. Xing, A. A. Belik, M. Arai, M. Kohno, T. Terashima, H. Kojitani, M. Akaogi, and K. Yamaura, Large anomalous Hall effect observed in the cubic-lattice antiferromagnet Mn$_3$Sb with kagome lattice, Phys. Rev. B {\bf 108}, 075140 (2023). 
\bibitem{Kotegawa_NbMnP}
H. Kotegawa, Y. Kuwata, V. T. N. Huyen, Y. Arai, H. Tou, M. Matsuda, K. Takeda, H. Sugawara, and M.-T. Suzuki, Large anomalous Hall effect and unusual domain switching in
an orthorhombic antiferromagnetic material NbMnP, npj Quantum Mater. {\bf 8}, 56 (2023). 
\bibitem{Kotegawa_TaMnP}
H. Kotegawa, A. Nakamura, V. T. N. Huyen, Y. Arai, H. Tou, H. Sugawara, J. Hayashi, K. Takeda, C. Tabata, K. Kaneko, K. Kodama, and M.-T. Suzuki, Large spontaneous Hall effect with flexible domain control in the antiferromagnetic material TaMnP, Phys. Rev. B {\bf 110}, 214417 (2024).
\bibitem{Kotegawa_216}
H. Kotegawa, H. Tanaka, Y. Takeuchi, H. Tou, H. Sugawara, J. Hayashi, and K. Takeda, Large Anomalous Hall Conductivity Derived from an $f$-Electron Collinear Antiferromagnetic Structure, Phys. Rev. Lett. {\bf 133}, 106301 (2024).


\bibitem{Qi}
J. Qi, W. Ren, C.-W. Wang, L. Zhang, C. Yu, Y. Zhuang, and Z. Zhang, Crystal and magnetic structures of Ce$_2$CuGe$_6$, J. Alloy. Comp. {\bf 805} 1260 (2019).
\bibitem{Sologub}
O. Sologub, K. Hiebl, P. Rogl, and O. I. Bodak, Formation, crystal chemistry and magnetism of compounds RE$_2$TGe$_6$, RE $\equiv$ rare earth, T $\equiv$ Pd, Pt, Cu, Ag and Au, J. Alloy. Comp. {\bf 227} 37 (1995).
\bibitem{Penc}
B. Penc, S. Baran, A. Hoser, J. Przewo\'{z}nik, A. Szytu{\l}a, Magnetic properties and magnetic structures of R$_2$PdGe$_6$ (R = Pr, Nd, Gd-Er) and R$_2$PtGe$_6$ (R = Tb, Ho, Er), J. Mag. Mat. Mat. {\bf 514}, 167152 (2020). 
\bibitem{Nakashima}
M. Nakashima, T. Kawai, T. Shimoda, T. Takeuchi, T. Yoneyama, T. D. Matsuda, Y. Haga, K. Shimizu, M. Hedo, Y. Uwatoko, and R. Settai, Y. \={O}nuki, Single crystal growth and pressure effect of an antiferromagnet Ce$_2$CuGe$_6$, Physica B {\bf 403},
789 (2008).
\bibitem{kaneko}
K. Kaneko, C. Tabata, M. Hagihala, H. Yamauchi, Y. Oba, T. Kumada, M. Kubota, Y. Kojima, N. Nabatame, M. Sasaki, Y. Shimojo, K. Kodama, and T. Osakabe, New standard for low temperature sample environment at JAEA/JRR-3, JPS Conf. Proc. {\bf 41}, 011015 (2024).
\bibitem{Strydom}
A. M. Strydom, A. V. Gribanov, Y. D. Seropegin, R. Wawryk, R. Tro\'c, Magnetic ordering and metamagnetism in Ce$_2$TGe$_6$ (T = Pd, Pt), J. Mag. Mat. Mat. {\bf 283}, 181 (2004). 
\bibitem{VESTA}
K. Momma and F. Izumi, Commission on Crystallogr. Comput., IUCr Newslett., {\bf 7}, 106 (2006).
\bibitem{Yaguchi}
T. Yaguchi, M. Nakashima, and Y. Amako, Single crystal growth and magnetic properties of $R_2$$T$Ge$_6$ ($R=$ Ce, Pr $T=$ Cu, Pd), JPS Conf. Proc. {\bf 30}, 011111 (2020).


\bibitem{Tian}
Y. Tian, L. Ye, and X. Jin, Proper Scaling of the Anomalous Hall Effect, Phys. Rev. Lett. {\bf 103}, 087206 (2009).
\bibitem{Onoda2008}
S. Onoda, N. Sugimoto, and N. Nagaosa, Quantum transport theory of anomalous electric, thermoelectric, and thermal Hall effects
in ferromagnets, Phys. Rev. B {\bf 77}, 165103 (2008).
\bibitem{Yang}
H.-Y. Yang, B. Singh, J. Gaudet, B. Lu, S.-M. Huang, G. Xu, Y. Zhao, C.-Y. Huang, W.-C. Chiu, B. Wang, F. Bahrami, B. Xu, J. Franklin, I. Sochnikov, D. E. Graf, C. M. Hoffman, H. Lin, D. H. Torchinsky, C. L. Broholm, A. Bansil, and F. Tafti, Noncollinear ferromagnetic Weyl semimetal with anisotropic anomalous Hall effect, Phys. Rev. B {\bf 103}, 115143 (2021).
\bibitem{Sakhya}
A. P. Sakhya, C.-Y. Huang, G. Dhakal, X.-J. Gao, S. Regmi, B. Wang, W. Wen, R.-H. He, X. Yao, R. Smith, M. Sprague, S. Gao, B. Singh, H. Lin, S.-Y. Xu, F. Tafti, A. Bansil, and M. Neupane, Observation of Fermi arcs and Weyl nodes in a noncentrosymmetric magnetic Weyl semimetal, Phys. Rev. Materials {\bf 7}, L051202 (2023).



\end{thebibliography}
\end{document}